\documentclass[preprint,onecolumn,nofootinbib]{revtex4}
\pdfoutput=1
\usepackage[colorlinks=true,linkcolor=blue,urlcolor=blue,filecolor=black,citecolor=red,
pdfstartview=FitV,pdftitle={},pdfsubject={},pdfkeywords={},pdfpagemode=None,bookmarksopen=true]{hyperref}
\usepackage{graphicx}
\usepackage{amsmath}
\usepackage{amsfonts,mathrsfs}
\usepackage{amssymb,ulem}
\usepackage{color}%
\usepackage{CJK}
\usepackage{dcolumn}
\newcommand{\f}{\begin{equation}}
\newcommand{\ff}{\end{equation}}
\newcommand{\fa}{\begin{eqnarray}}
\newcommand{\ffa}{\end{eqnarray}}

\begin{document}

\title{Dynamic Properties of Two-Dimensional Latticed Holographic System}
\author{Peng Liu $^{1}$}
\thanks{phylp@email.jnu.edu.cn}
\author{Jian-Pin Wu $^{2}$}
\thanks{jianpinwu@yzu.edu.cn}
\affiliation{
  $^{1}$ Department of Physics and Siyuan Laboratory, Jinan University, Guangzhou 510632, P.R. China\ \\
  $^{2}$ Center for Gravitation and Cosmology, College of Physical Science and Technology, Yangzhou University, Yangzhou 225009, China\ \\
}
\begin{abstract}

  We study the anisotropic properties of dynamical quantities: direct current (DC) conductivity, butterfly velocity, and charge diffusion. The anisotropy plays a crucial role in determining the phase structure of the two-lattice system. Even a small deviation from isotropy can lead to distinct phase structures, as well as the IR fixed points of our holographic systems. In particular, for anisotropic cases, the most important property is that the IR fixed point can be non-AdS$_2 \times \mathbb R^2$ even for metallic phases. As that of a one-lattice system, the butterfly velocity and the charge diffusion can also diagnose the quantum phase transition (QPT) in this two-dimensional anisotropic latticed system.

\end{abstract}
\maketitle
\tableofcontents
\section{Introduction}

Anisotropy is universal and results in many rich phenomena in nature, such as magnetic systems, latticed systems, and so on \cite{Glotzer:2007}. In some strongly correlated systems, the anisotropy is associated with many dynamical measures, for example, the transport properties. The effects of anisotropy on dynamical measures for strongly correlated systems are useful for practical u Strongly correlated systems, however, are long-standing hard problems in physics. Gauge/gravity duality can bring novel insight and understanding into the strongly correlated systems and offer a good platform to study the anisotropy of dynamical quantities in strongly correlated systems.

Gauge/gravity duality has been proved powerful tools to study strongly correlated systems and dynamical properties \cite{Maldacena:1997re,Gubser:1998bc,Witten:1998qj,Aharony:1999ti,Hartnoll:2016apf}. Anisotropy is also ubiquitous in holographic systems, such as systems with lattices, anisotropic axions, massive gravity, and so on \cite{Ling:2015ghh,Fang:2014jka,Arefeva:2018hyo,Jeong:2017rxg,Cremonini:2014pca,Grandi:2021bsp,Baggioli:2020qdg}. All these models realize the anisotropy by explicitly breaking the isotropic symmetry. The anisotropy can also be introduced by spontaneous symmetry breaking, such as holographic charge density wave models \cite{Donos:2013gda,Ling:2014saa}.

Due to the breaking of translational symmetry from lattices, the DC conductivity is finite. In \cite{Donos:2014uba}, a novel technics to calculate the DC conductivity from lattice systems has been developed. Using these technics, a simple analytic expression for the DC conductivity of a large class of holographic lattice models can be obtained, see \cite{Blake:2014yla,Donos:2014cya,Zhou:2015dha} and therein. In particular, in terms of the relation between the DC conductivity and the temperature, it is convenient to work out the (metal or insulating) phase of the system, see for instances \cite{Donos:2014oha,Ling:2016dck,Wu:2018pig,Ling:2015dma,Ling:2015epa,Ling:2015exa,Ling:2016wyr,Ling:2016ibq,Baggioli:2014roa,Baggioli:2016oqk,Baggioli:2016oju,An:2020tkn}.

This leads to three interesting possibilities: the system is a metallic phase in any direction, the system is a metallic phase in one direction and an insulating phase in another direction, and the system is an insulating phase in any direction. We are particularly concerned about the quantum phase transitions corresponding to these three different states and their relationship with anisotropy.

The butterfly effect ubiquitously exists in holographic systems and is depicted by a shockwave solution of a black hole near the horizon \cite{Roberts:2016wdl,Shenker:2013pqa,Shenker:2013yza,Shenker:2014cwa,Blake:2016wvh}. Since the Lyapunov exponent in holographic system saturates the bound of chaos \cite{Maldacena:2015waa}, the butterfly velocity is an important dynamical quantity to signal some characteristics of holographic system. In particular, the butterfly velocity is proposed as the characteristic quantity of holographic system \cite{Blake:2016wvh,Blake:2016sud,Lucas:2016yfl,Hartnoll:2014lpa}. In addition, it has been found that the butterfly velocity can characterize the phase transition \cite{Ling:2016ibq,Ling:2016wuy,Baggioli:2018afg}.

In addition, the charge diffusion, as an important dynamic property, can play a crucial role in explaining the strange metals in incoherent limit. \cite{Hartnoll:2014lpa} suggested that the universal strange metals should be related to the ubiquitous charge diffusion, and proposed a bound for the diffusion constant. In \cite{Blake:2016wvh}, the butterfly velocity in strongly correlated systems is considered as the characteristic velocity to saturate the diffusion constant bound. Therefore, it is worth exploring the relationship between charge diffusion and quantum phase transition.

In this paper, we study DC conductivity, butterfly velocity and the charge diffusion over an anisotropic background from Q-lattice. Most of the previous work only studied anisotropy with the one-dimensional lattice. Here, we shall derive the expressions of the DC conductivity and the butterfly velocity in the holographic system with the two-dimensional lattice. And it would be crucial to check if those behaviors in the holographic system with one-dimensional lattice are still valid for two-dimensional cases. In this paper, we shall especially examine that how MIT is affected by the anisotropy and to what extent the characterization of the phase transition by butterfly velocity is universal. We organize this paper into three parts: we introduce the 2-dimensional Q-lattice model in \ref{qla-ani}; then we study the anisotropic dynamical quantities, including the DC conductivity and the butterfly velocity, and the related properties, in \ref{sec:dynamical}; we study the charge diffusion in \ref{sec:charge};  we conclude and discuss in \ref{sec:discussion}.

\section{Holographic Q-lattice model}\label{qla-ani}
The holographic Q-lattice model is a concise realization of the periodic structure. Previous holographic lattice models, such as the ionic lattices model and the scalar lattices model, introduce spatially periodic structures on scalar fields or the chemical potential (see \cite{Ling:2015ghh} for a recent review). The resultant equations of motion are a set of highly nonlinear partial differential equations, which pose a challenge for numerical solutions. By contrast, the Q-lattice model introduces a complex scalar field, which results in only ordinary differential equations. Therefore, the Q-lattice model is an easier realization of lattice structures. The Q-lattice model is useful in modeling the Mott insulator and MIT \cite{Ling:2015exa,Ling:2015epa}.

The Lagrangian of the Q-lattice model is \cite{Donos:2013eha,Donos:2014uba,Ling:2014laa,Ling:2015dma},
\begin{equation}\label{actionq}
  \mathcal L = R + 6 - \frac{1}{2}F^{ab} F_{ab} + \partial_a \Phi_1^\dagger \partial^a \Phi_1 + \partial_a \Phi_2^\dagger \partial^a \Phi_2 +  m_{1}^2 |\Phi_{1}|^2 + m_{2}^2 |\Phi_{2}|^2.
\end{equation}
System \eqref{actionq} can be solved with ansatz,
\begin{equation}\label{ansatz}
  \begin{aligned}
    d s ^ { 2 } & = \frac { 1 } { z ^ { 2 } } \left[ - ( 1 - z ) p ( z ) U d t ^ { 2 } + \frac { d z ^ { 2 } } { ( 1 - z ) p ( z ) U } + V _ { 1 } d x ^ { 2 } + V _ { 2 } d y ^ { 2 } \right],   \\
    A           & = \mu ( 1 - z ) a d t, \qquad \Phi_{1}  = e ^ { i k_{1} x } z ^ { 3 - \Delta_{1} } \phi_{1}, \qquad \Phi_{2}  = e ^ { i k_{2} y } z ^ { 3 - \Delta_{2} } \phi_{2}
  \end{aligned}
\end{equation}
where $p ( z ) = 1 + z + z ^ { 2 } - \mu ^ { 2 } z ^ { 3 } / 2 $ and $ \Delta_{1,2} = 3 / 2 \pm \left( 9 / 4 + m_{1,2} ^ { 2 } \right) ^ { 1 / 2 }$. For concreteness, we set $m_{1}^{2} = m_{2}^{2}=m^2 = -2$, therefore we have $\Delta_{1} = \Delta_{2}=\Delta = 2$.
The case $\Delta_1\neq \Delta_2$ means that the lattice in different directions may have different relevance or irrelevance, which results in anisotropic systems such as \cite{hosseini:2007}.
The horizon and the boundary locate at $z=1$ and $z=0$ respectively. The $A_{a}$ is the Maxwell field, and $\phi_{1,2}$ is the complex scalar field mimicking the lattice structures. Consequently, the functions to solve are $\left(a,\phi,U,V_{1},V_{2}\right)$.

To solve the system \eqref{actionq}, we need to specify the boundary conditions and system parameters. We set $a(0)=1$, then $A_{t}(0) = \mu$ becomes the chemical potential of the dual system. The boundary condition $\lambda_{1,2}= \phi_{1,2}(0)$ is the strength of the lattice deformation, and $k_{1,2}$ is the wave vector of the periodic structure. The asymptotic AdS$_{4}$ requires that $U(0)=1, V_{1}(0)=1,V_{2}(0)=1$. Other boundary conditions at the horizon $(z=1)$ can be fixed by regularity. The Hawking temperature reads $T = \left( 6 - \mu ^ { 2 } \right) U ( 1 ) / ( 8 \pi )$. The black brane solutions can be categorized by $5$ dimensionless parameters $\left(\tilde T,\tilde\lambda_1,\tilde\lambda_2,\tilde k_1,\tilde k_2\right)\equiv\left(\frac{T}{\mu},\frac{\lambda_1}{\mu},\frac{ \lambda_2}{\mu},\frac{k_1}{\mu},\frac{k_2}{\mu}\right)$, where we adopt the chemical potential $\mu$ as the scaling unit. Furthermore, due to the scaling symmetry of the system, only scale invariant physical quantities will be discussed in this paper. This means that any physical quantity with scaling dimension $\Delta$ will be divided by $\mu^{\Delta}$ to cancel its scaling dimension.

\section{Anisotropy effect on dynamical quantities}\label{sec:dynamical}

\subsection{The DC conductivity}

The DC conductivity $\sigma_{\text{DC}}$ can be obtained with horizon data. For an anisotropic background, the DC conductivity is anisotropic as well. Previous studies on anisotropy focused only on two distinct directions. To reveal more detailed anisotropic behavior of the DC conductivity, we study the DC conductivity along an arbitrary direction, i.e., the anisotropic DC conductivity.

\begin{figure}[htbp]
  \centering
  \includegraphics[width=0.4\textwidth]{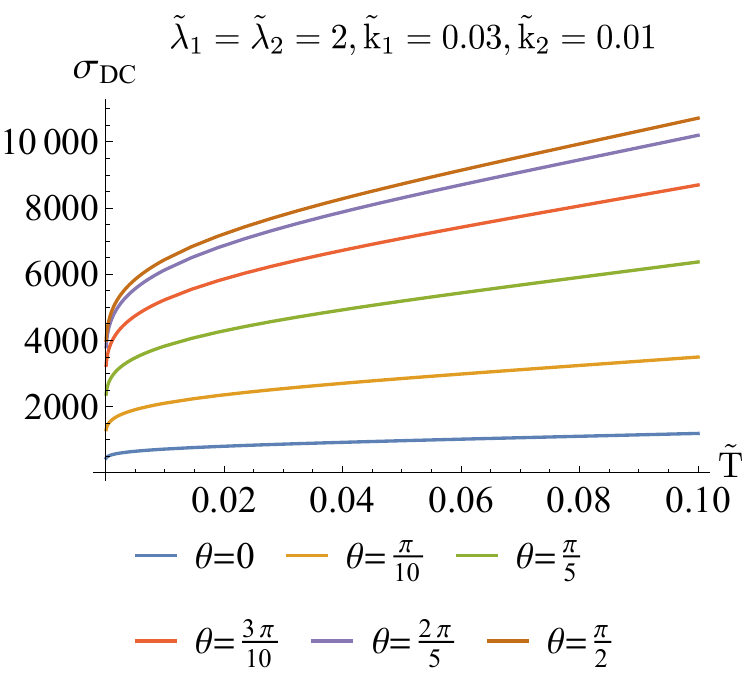}\ \hspace{0.5cm}
  \includegraphics[width=0.4\textwidth]{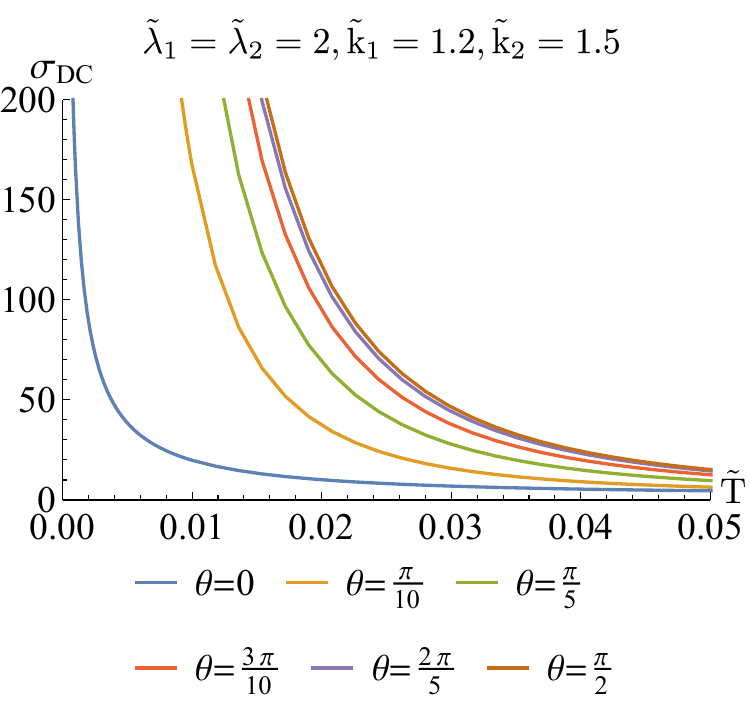}\ \\
  \includegraphics[width=0.4\textwidth]{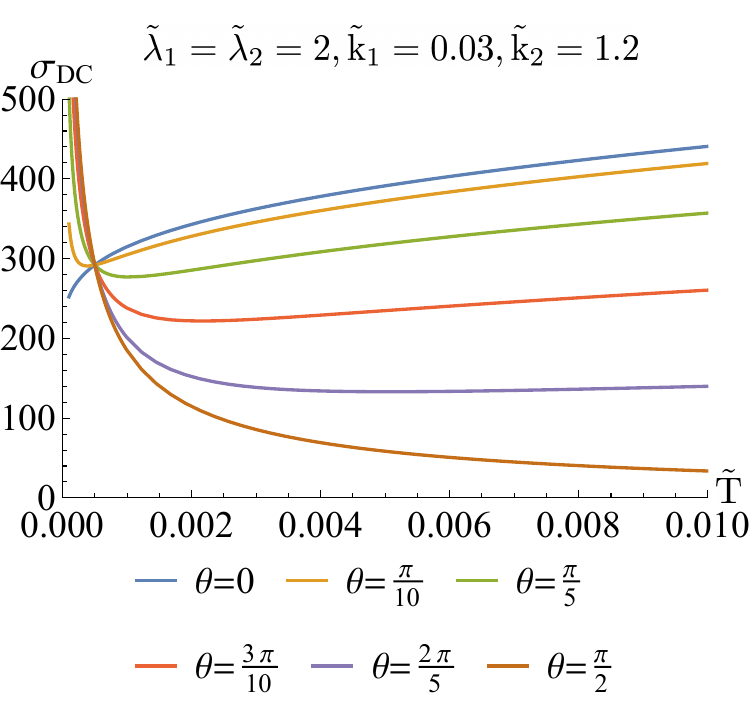}\ \hspace{0.5cm}
  \includegraphics[width=0.4\textwidth]{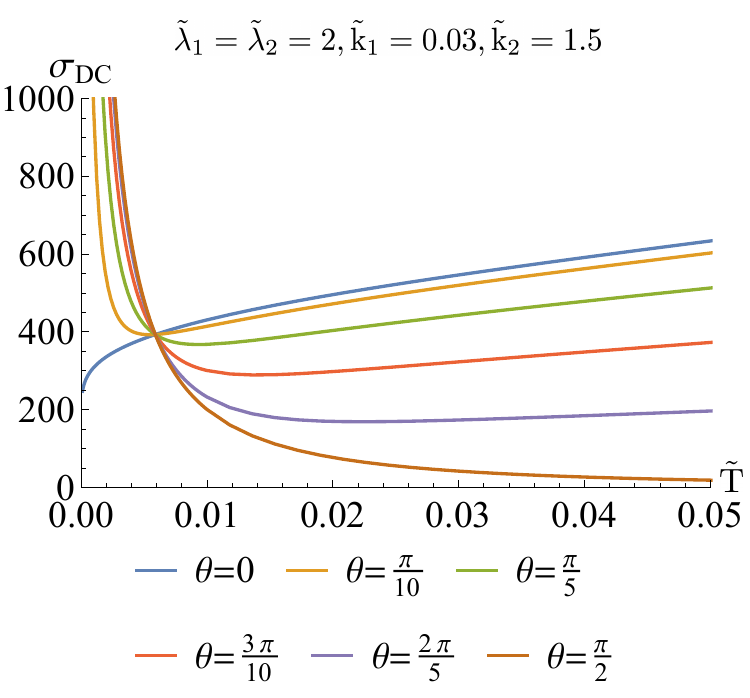}\ \\
  \caption{$\sigma_{DC}$ as the function of $\tilde T$ for different wave number and angle $\theta$. The intersection point is the place where $\sigma_{x} = \sigma_{y}$.}
  \label{fig:dcvst}
\end{figure}

In Fig. \ref{fig:dcvst}, we can find that the upper left plot and the upper right plot correspond to the insulating phase and metallic phases, respectively. For the lower row plots, the system is insulating in $x$-direction, while metallic in $y$-direction. Also, the intersection point in both plots of the lower row suggests that $\sigma_{x} = \sigma_{y}$ at certain temperature, which results in isotropic $\sigma$.

\begin{figure}[htbp]
  \centering
  \includegraphics[width=0.5\textwidth]{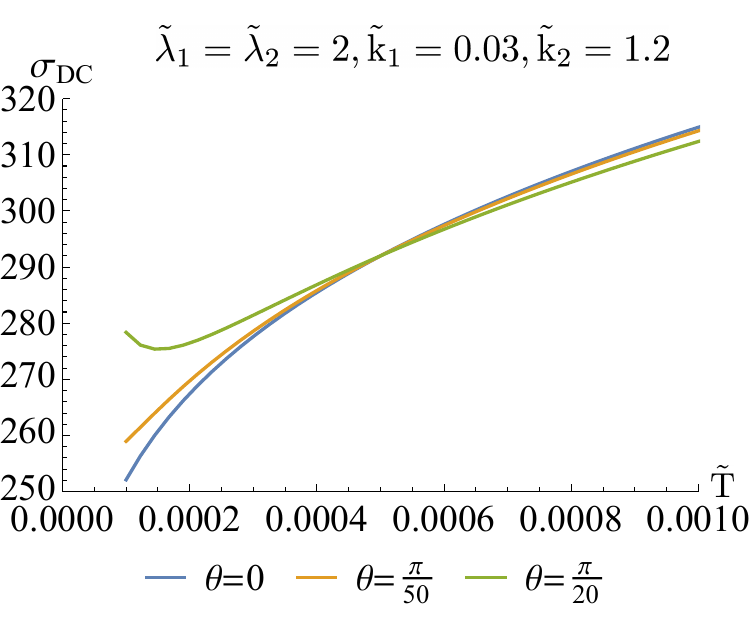}
  \caption{$\sigma_{DC}$ as the function of $\tilde T$ for $\tilde\lambda_1=\tilde\lambda_2=2$, $\tilde k_1=0.03$ and $\tilde k_2=1.2$ in the region of small $\theta$.}
  \label{fig:dcvstv2}
\end{figure}

To obtain the anisotropic DC conductivity, we study the linear responses behaviors of the current
\begin{equation}\label{eq:anisores}
  J_i = R_{ij} E_j,
\end{equation}
where $J_i$, $E_j$, $R_{ij}$ are current, electric field and response matrix, respectively. Following the method in \cite{Donos:2014uba} (also see \cite{Blake:2014yla,Ling:2016dck}), we can calculate the DC conductivity $\sigma_{x\mathrm{DC}}, \sigma_{y\mathrm{DC}}$ along the $x,y$ direction, respectively
\begin{eqnarray}\label{eq:dcxy}
  \sigma_{x\mathrm{DC}}&=R_{11}=\left.\left(\sqrt{\frac{V_{2}}{V_{1}}}+\frac{\mu^{2} a^{2} \sqrt{V_{1} V_{2}}}{k_1^{2} \phi_1^{2}}\right)\right|_{z=1},\\
  \sigma_{y\mathrm{DC}}&=R_{22}=\left.\left(\sqrt{\frac{V_{1}}{V_{2}}}+\frac{\mu^{2} a^{2} \sqrt{V_{1} V_{2}}}{k_2^{2} \phi_2^{2}}\right)\right|_{z=1},
\end{eqnarray}
with the off-diagonal components $R_{12} = R_{21} = 0$. Therefore the angle-dependent DC conductivity reads,
\begin{equation}\label{eq:dcangle}
  \begin{aligned}
    \sigma_{\text{DC}}(\theta) & = |J^a|/|E^a| =
    \sqrt{J^a J_a / (E^b E_b)} = \sqrt{R_{ij}E_j R_{ik} E_k / (E_s E_s)}                                                                                                           \\
                               & = \sqrt{(R_{11}^2 E_1^2 + R_{22}^2 E^2_2)/ (E_1^2 + E_2^2)} = \sqrt{\sigma_{x\mathrm{DC}}^2 \cos^2\theta + \sigma_{y\mathrm{DC}}^2\sin^2 \theta}.
  \end{aligned}
\end{equation}
Once the DC conductivity is at hand, we can define the metallic phase as $\sigma'_{DC}(\tilde T)<0$ and the insulating phase as $\sigma'_{DC}(\tilde T)>0$. From this definition, we can see that as long as the system is in the metallic phase in one direction, the system is always in the metallic phase in any direction except the direction perpendicular to it where the system is in the insulating phase (See Fig. \ref{fig:dcvstv2}). Moreover, the insulating phase in any direction requires the system to be insulating phase along the direction of two lattices.

Now we study the DC conductivity as the function of the temperature $\tilde T$ for different angles $\theta$.
For simplicity, we take $\tilde\lambda_1=\tilde\lambda_2=2$ here and the anisotropy is realized by setting $\tilde k_1\neq \tilde k_2$.
We find that when the system is insulating or metallic along both $x$ and $y$ directions,
the system is also insulating or metallic for arbitrary angles (see the plots above in Fig. \ref{fig:dcvst}).
However, if the system is insulating along $x$ direction but metallic along $y$ direction,
some interesting phenomena emerge. We summarize the observations as follows.
\begin{itemize}
  \item When $\theta$ is small, the system is insulating as expected (see the plots in Fig. \ref{fig:dcvst} and Fig. \ref{fig:dcvstv2}).
  \item As $\theta$ increases, it is observed that the DC conductivity goes down as the temperature decreases, which seems to be an insulating state, but after reaching its minimum it turns to rise with the decrease of the temperature, which is metallic behavior. A similar phenomenon is also observed in the holographic doped Mott system studied in \cite{Ling:2015epa}.
  \item When $\theta$ becomes larger, the system becomes metallic, which is also as expected (see the plots in Fig. \ref{fig:dcvst}).
\end{itemize}
The above phenomena match the analysis on the metallic in an arbitrary direction in the previous paragraph.

The critical angle at which the system passes from metallic phase to insulating phase can be also worked out. By $d\sigma_{DC}/d\tilde T=0$, we obtain the critical angle $\theta_c$ satisfying
\fa
\label{thetac}
\tan^2(\theta_c)=-\frac{\sigma_{xDC}\sigma'_{xDC}}{\sigma_{yDC}\sigma'_{yDC}}\,,
\ffa
where the prime denotes the derivative with respect to the temperature. For $\tilde\lambda_1=\tilde\lambda_2=2$, $\tilde k_1=0.03$ and $\tilde k_2=1.2$, we work out the critical angle $\theta_c\simeq 0.116$ at the temperature $\tilde T=10^{-4}$.

\begin{figure}[htbp]
  \centering
  \includegraphics[width=0.5\textwidth]{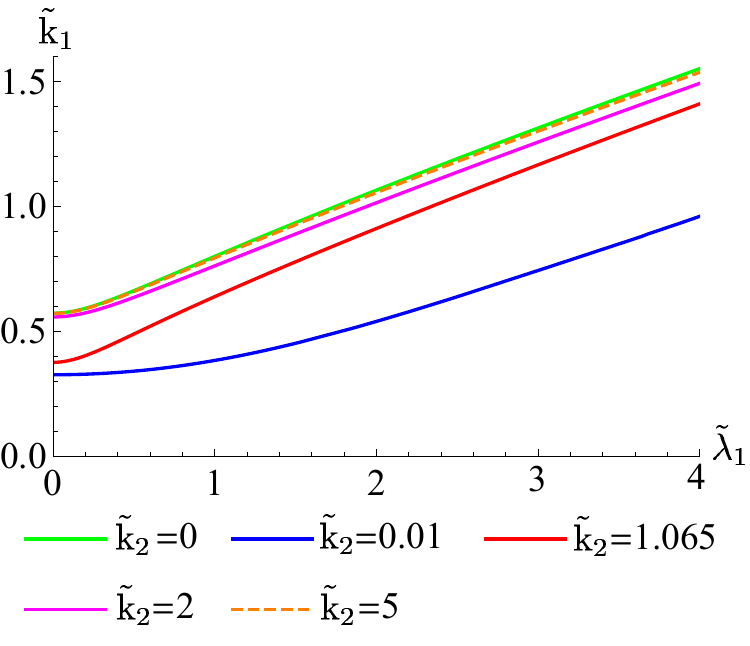}
  \caption{MIT phase diagram ($\tilde\lambda_1$, $\tilde k_1$) for $\tilde\lambda_2=2$ and different $\tilde k_2$ at $\tilde T=0.001$. The phase above the critical line is the metallic phase and the one below the critical line is the insulating phase. For the green curve, it is the $\tilde\lambda_2 = 0$ case, where the second lattice disappears.}
  \label{fig:mitpd}
\end{figure}

Compared with the one-dimensional lattice system, the extra lattice can not only realize the phase transition in the other direction, but also affect the phase of the previous lattice system. Next, we shall study the MIT phase diagram. In our previous works \cite{Ling:2015dma}, MIT phase diagram ($\tilde\lambda_1$, $\tilde k_1$) for $\tilde k_2=0$ has been worked out. Here, we want to study the effect from the lattice along $y$ direction on the phase diagram ($\tilde\lambda_1$, $\tilde k_1$). So we plot the phase diagram ($\tilde\lambda_1$, $\tilde k_1$) for different $\tilde k_2$ at $\tilde T=0.001$ (Fig. \ref{fig:mitpd}). The properties are summarized as follows.
\begin{itemize}
  \item For fixed $\tilde\lambda_1$, the state of system changes from metal to insulator as $\tilde k_1$ decreases for different $\tilde k_2$.
        It is similar to that for $\tilde k_2=0$, which has been studied in \cite{Ling:2015dma}.
  \item For $\tilde k_2\neq 0$, with the increase of $\tilde k_2$, the critical line goes up,
        which means that the phase transition from metal to insulator happen at larger $\tilde k_1$ ($\tilde\lambda_1$ is fixed).
        It is because the phase transition from insulator to metal happens along $y$ direction as $\tilde k_2$ increases, and when $\tilde k_2$ is further tuned larger,
        the system along $y$ direction exhibits more obvious metallic behavior.
  \item $\tilde k_2=0$ sets the upper bound of the critical line in phase diagram ($\tilde\lambda_1$, $\tilde k_1$).
        As $\tilde k_2$ increases, this critical line goes up and approaches this upper bound.
\end{itemize}

To understand the above phenomena, we implement the mode analysis for AdS$_2 \times \mathbb R^2$ on this two-lattice system.
\begin{equation}
  \label{eq:modeset}
  \begin{aligned}
    U & = 6 r ^ { 2 } \left( 1 + u _ { 1 } r ^ { \delta } \right) , \quad V _ { 1 } = v _ { 10 } \left( 1 + v _ { 11 } r ^ { \delta } \right) , \quad V _ { 2 } = v _ { 20 } \left( 1 + v _ { 21 } r ^ { \delta } \right), \\
    a & = 2 \sqrt { 3 } r \left( 1 + a _ { 1 } r ^ { \delta } \right) , \quad \phi_1 = e ^ { i \tilde k_1 x _ { 1 } } \varphi _ { 1 } r ^ { \delta } ,\quad \phi_2 = e ^ { i \tilde k_2 x _ { 1 } } \varphi _ { 2 } r ^ { \delta }.
  \end{aligned}
\end{equation}
We find that the mode expansions of these two lattice fields are independent from each other. Specifically, we find the scaling dimensions of the two lattice fields read
\begin{equation}\label{eq:moderes_1}
  \delta _ { \varphi_1 } = - \frac { 1 } { 2 } + \frac { 1 } { 2 \sqrt { 3 } } \sqrt { 3 + 2 m ^ { 2 } + 2 e ^ { - 2 v _ { 10 } } \tilde k_1 ^ { 2 } }, \quad
  \delta _ { \varphi_2 } = - \frac { 1 } { 2 } + \frac { 1 } { 2 \sqrt { 3 } } \sqrt { 3 + 2 m ^ { 2 } + 2 e ^ { - 2 v _ { 20 } } \tilde k_2 ^ { 2 } }.
\end{equation}
Therefore, the lattice will deform away the AdS$_2\times \mathbb R^2$ as long as either of,
\begin{equation}
  \label{eq:moderes_2}
  \left( e ^ { - v _ { 10 } } \tilde k_1 \right) ^ { 2 } < - m ^ { 2 },\quad \left( e ^ { - v _ { 20 } } \tilde k_2 \right) ^ { 2 } < - m ^ { 2 },
\end{equation}
is satisfied.

From the \eqref{eq:moderes_2} we can see that, for small values of $\tilde k_2$ the IR fixed point is no longer AdS$_2 \times \mathbb R^2$, and hence it will lay significant effects on the phase structure of the first lattice. However, we also find that when $\tilde k_2$ is large, the effect of the second lattice on the first lattice disappears.  This can also be seen from \eqref{eq:moderes_2}. Obviously, when $\tilde k_2$ is large, the second lattice is irrelevant. As a result, in the action \eqref{actionq}, the terms involving $\Phi_2$ in the near horizon region will vanish, so it will not have a significant impact on the phase structure along the first lattice. In addition, from the dual point of view, this is a very physical expected result. In the dual picture, large $\tilde k_2$ means small lattice spacing. The electron can easily move between different sites{. Therefore, the electrons will not be localized along the second lattice, instead, they behave more like free electrons. Therefore, when $\tilde k_2$ is large, the degree of freedom of the second lattice becomes redundant, it will not have a significant impact on the phase diagram along the first lattice.

It is worth noting that although the IR fixed points in many models can be obtained by analytical treatment, and verified by numerical treatment \cite{Donos:2014uba}. However, the IR fixed points of the model in the current paper still ask for further investigation \cite{Donos:2013eha}. Nevertheless, the butterfly velocity we discuss next will show that the metallic phase and the insulating phase do correspond to different fixed points.

\subsection{The butterfly velocity and QPT}

In this section, we study the anisotropic butterfly velocity. In our previous works \cite{Ling:2016ibq,Ling:2016wuy}, the anisotropic butterfly velocity has been explored in special case of $\tilde\lambda_2=0$ and $\tilde k_2=0$. Here, we shall turn on the lattice along $y$ direction and see how it affects the butterfly velocity.

In original coordinate $(t,z,x,y)$ the butterfly velocity is anisotropic. Since the butterfly velocity only depends on the horizon data, therefore, one may implement a convenient coordinate transformation $ \tilde y = \sqrt{\frac{V_{yy}}{V_{xx}}} y $, and in the new coordinate $(t,z,x,\tilde y)$, the butterfly velocity $v_B$ is isotropic again. When going back to the original coordinate, the angle-dependent butterfly velocity $\mathfrak{v}_B(\theta)$ can be obtained as,
\begin{equation}\label{eq:anisovb}
  \mathfrak{v}_{B}(\theta) =  v_{B} \sqrt{\cos ^2(\theta )+\frac{V_{xx} \sin ^2(\theta )}{V_{yy}}}.
\end{equation}

\begin{figure}[htbp]
  \centering
  \includegraphics[width=0.45\textwidth]{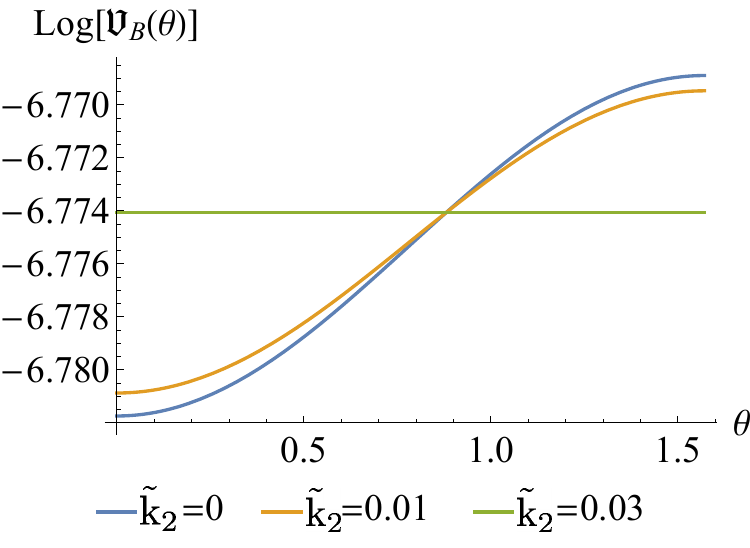}\
  \includegraphics[width=0.45\textwidth]{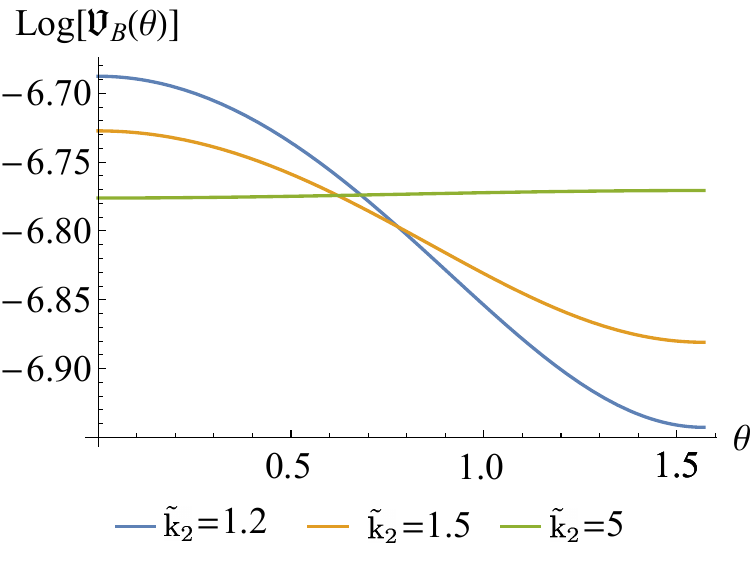}\ \\
  \caption{$\mathfrak{v}_{B}(\theta)$ as the function at $\tilde k_1=0.03$ and $\tilde T=0.0001$ with different $\tilde k_2$.}
  \label{fig:vbvstheta}
\end{figure}

From Eq. \eqref{eq:anisovb}, we can infer that the period of $\mathfrak{v}_{B}(\theta)$ is $\pi$ and $\mathfrak{v}_{B}(\pi/2-\theta)=\mathfrak{v}_{B}(\pi/2+\theta)$ and so we shall only focus on the angle range $\theta\in[0,\pi/2]$ here. We show $\mathfrak{v}_{B}(\theta)$ as the function of $\theta$ at $\tilde\lambda_1=\tilde\lambda_2=2$, $\tilde k_1=0.03$ with different $\tilde k_2$ in Fig. \ref{fig:vbvstheta}, from which we can see that for $\tilde k_2<\tilde k_1$, $\mathfrak{v}_{B}(\theta)$ monotonically increases as $\theta$ increases (left plot in Fig. \ref{fig:vbvstheta}), but for $\tilde k_2>\tilde k_1$, it monotonically decreases as $\theta$ increase (right plot in Fig. \ref{fig:vbvstheta})\footnote{Without loss of generality, we shall fix $\tilde\lambda_1=\tilde\lambda_2=2$ in the subsequent study and only change $\tilde k_1$ and $\tilde k_2$ to implement the anisotropy geometry.}. Therefore, we can conclude that the lattice suppresses the butterfly velocity. It is no question that when $\tilde k_1=\tilde k_2$, $\mathfrak{v}_{B}(\theta)$ is a constant independent of $\theta$ (green line in left plot in Fig. \ref{fig:vbvstheta}). In addition, when $\tilde k_2\gg \tilde k_1$, $\mathfrak{v}_{B}(\theta)$ also is approximately a constant (green line in right plot in Fig. \ref{fig:vbvstheta}).

\begin{figure}[htbp]
  \centering
  \includegraphics[width=0.45\textwidth]{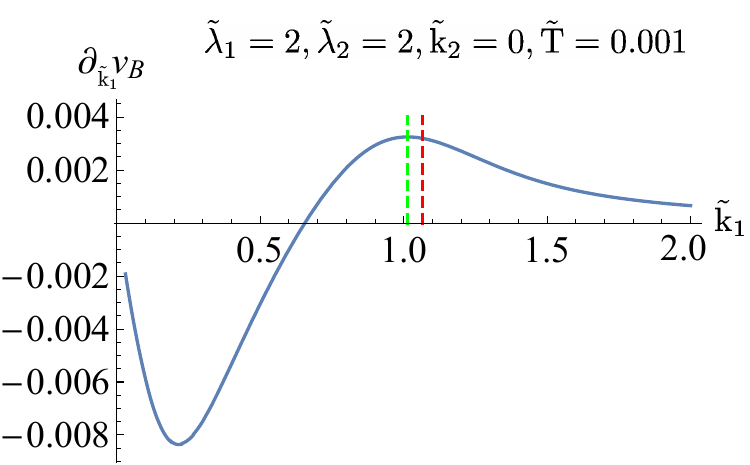}\
  \includegraphics[width=0.45\textwidth]{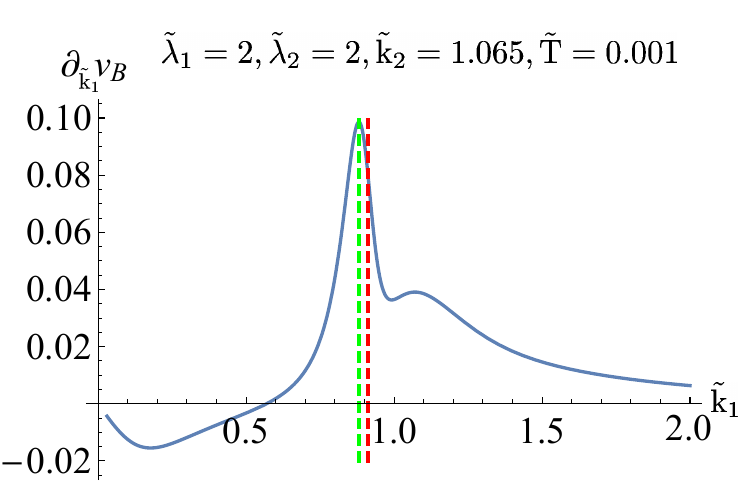}\ \\
  \includegraphics[width=0.45\textwidth]{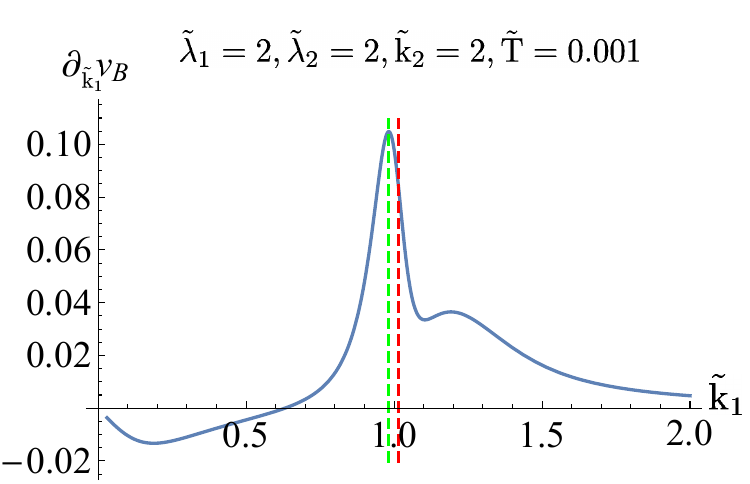}\
  \includegraphics[width=0.45\textwidth]{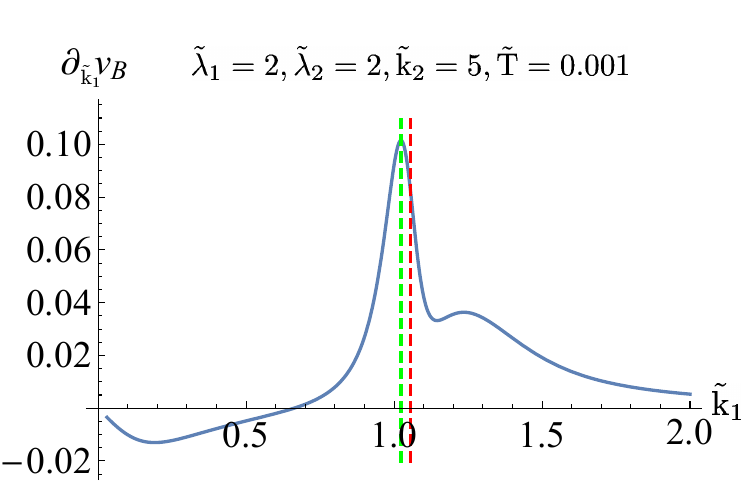}\ \\
  \caption{$\partial_{\tilde k_1}v_B$ as the function of $\tilde k_1$ at $\tilde T=0.001$. The red dashed line represents the position of the critical point while the green dashed line denotes the position of the local maximum of $\partial_{\tilde k_1}v_B$.}
  \label{fig:pdvbvsk}
\end{figure}

We also study the behavior of butterfly velocity $v_B$ along $x$ direction near the QPT. Fig. \ref{fig:pdvbvsk} shows $\partial_{\tilde k_1}v_B$ as the function of $\tilde k_1$ at $\tilde T=0.001$ with different $\tilde k_2$. We find that even when the lattice effect along $y$ direction is introduced, $v_B$ along $x$ direction can diagnose the QPT along $x$ direction. It indicates there is a rapid change of $v_B$ as well as the local extremes of $\partial_kv_B$ near the QCPs.

The essence of the above phenomenon is that the occurrence of quantum phase transition is usually accompanied by the change of IR fixed point. The power-law relationship between butterfly velocity and temperature at zero temperature limit is directly determined by IR fixed point. Therefore, the first derivative of the butterfly velocity with the system parameters will show an obvious peak behavior when it crosses the fixed point.

Notice that the butterfly velocity presents a discontinuity in the first-order derivative in the process of holographic superconductivity phase transition \cite{Ling:2016wuy}. It is worth noting that the phenomena in quantum phase transition are different from this, and the mechanisms behind them are also completely different. The superconductivity phase transition will be accompanied by condensation, and the background geometry of the system will certainly be modified. From the perturbation analysis, it can be seen that the order of the perturbation received by the metric is twice that of the condensation field. Therefore, it can be concluded that the butterfly velocity in \cite{Ling:2016wuy} has an obvious scaling law at the critical point, and its critical exponent is $1$. The MIT phase transition discussed in this paper is driven by the change of IR fixed point, and there is no emergence of the condensation. Therefore, the power-law dependence of butterfly velocity on temperature can well reflect this point.

Next, we want to study the behavior of $v_B$ in the zero-temperature limit.

\subsection{The scaling behavior of the butterfly}
\begin{figure}[htbp]
  \centering
  \includegraphics[width=0.7\textwidth]{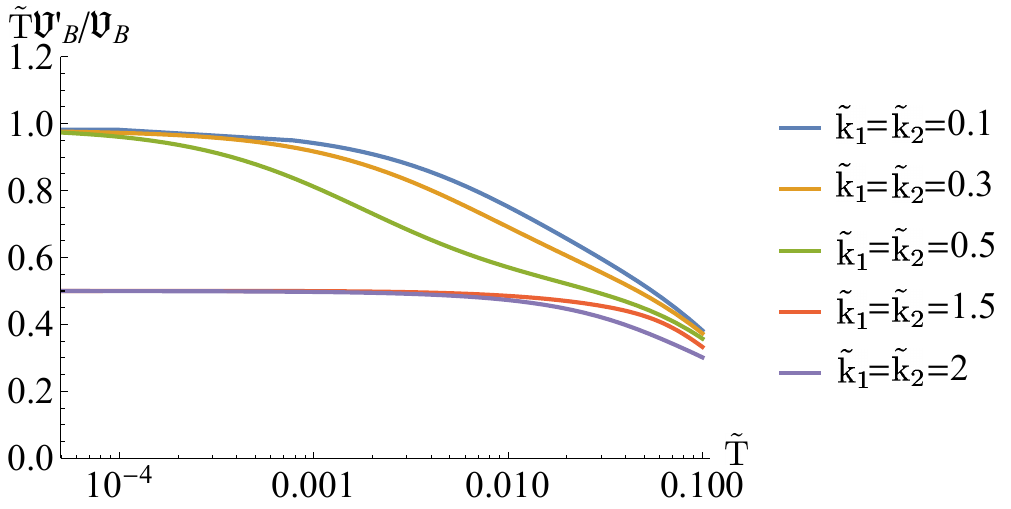}\ \\
  \caption{$\tilde T\mathfrak v_B'/\mathfrak v_B \,v.s.\, \tilde T$ for different phases over isotropy background ($\tilde k_1=\tilde k_2=1.5,\,2$ corresponds
    to metallic phases and $\tilde k_1=\tilde k_2=0.1,\,0.3,\,0.5$ corresponds to insulating phases).
  }
  \label{fig:vbvstvsiso}
\end{figure}

In Ref. \cite{Blake:2016sud}, they deduced that metallic phases in one-dimensional Q-lattice model always correspond to the well-known AdS$_2 \times \mathbb R^2$ IR geometry, on which $v_B \sim \tilde T^{1/2}$. And then, in our previous work \cite{Ling:2016ibq}, we find that for the one-dimensional Q-lattice model, the scaling of $v_B$ with temperature is $v_B\sim \tilde T^{\alpha}$ in both metallic and insulating phases, but the exponent $\alpha$ approaches $1$ (in fact it is slightly larger than $1$) in insulating phase, which is different from that in metallic phase. It indicates the theory flows to different IR fixed points for metallic and insulating phases, respectively.

We also assume that in two-dimensional Q-lattice model the scaling of $v_B$ follows the same behavior as $v_B\sim \tilde T^{\alpha}$. In addition, we also assume $V_{xx}/V_{yy}\sim \tilde T^\beta$. Then, $\mathfrak v_B(\theta)$ becomes,
\begin{equation}
  \label{eq:texp}
  \mathfrak v_B(\theta)\sim \tilde T^{\alpha } \sqrt{\cos ^2(\theta )+\sin ^2(\theta ) \tilde T^{\beta }}
\end{equation}
Therefore we have the temperature power exponent as
\begin{equation}
  \label{eq:texpdv0}
  \tilde T\mathfrak v_B'/\mathfrak v_B \sim \frac{2\alpha\cos^2(\theta)+(2\alpha+\beta)\sin^2(\theta)\tilde T^\beta}{2(\cos^2(\theta)+\sin^2(\theta)\tilde T^\beta)}\,,
\end{equation}
where $\prime$ denotes the derivative with respect to the temperature $\tilde T$. Explicitly, we can work out the temperature power exponent for some special $\theta$ as
\begin{equation}
  \label{eq:texpd}
  \tilde T\mathfrak v_B'/\mathfrak v_B \sim \left\{
  \begin{aligned}
     & \alpha                                           &  & \theta=0     \\
     & \alpha +\frac{\beta }{2}                         &  & \theta=\pi/2 \\
     & \alpha +\frac{\beta  \tilde T^{\beta }}{2 \tilde T^{\beta }+2} &  & \theta=\pi/4
  \end{aligned}
  \right.,
\end{equation}

Now, we begin to numerically study the behavior of $\mathfrak v_B$ in zero-temperature limit. And then we also try to understand the corresponding behavior based on the above assumption. Fig. \ref{fig:vbvstvsiso} exhibits $\tilde T\mathfrak v'_B/\mathfrak v_B$ as the function of $\tilde T$ for different phases over isotropy background. We clearly see that for metallic phases, $\alpha=1/2$. As expected it is the same as the metallic phase of the one-dimensional Q-lattice model. This suggests that the metallic phases for the isotropic 2-lattice system are essentially the same as that of the 1-lattice system. This is as expected since adding an extra copy in a perpendicular direction shall not change the IR geometry too much. However, for insulating phases, $\alpha$ approaches $1$ but is slightly smaller than $1$. It indicates that the insulating phase over isotropy two-dimensional Q-lattice model flows a novel IR fixed point, which is different from the well-known AdS$_2 \times \mathbb R^2$ IR geometry as expected, is also different from that of one-dimensional Q-lattice model where the $\alpha$ approaches a fixed value slightly larger than unity. Given the mild difference for insulating phases, the insulating phases indeed converge to certain fixed values.

\begin{figure}[htbp]
  \centering
  \includegraphics[width=0.7\textwidth]{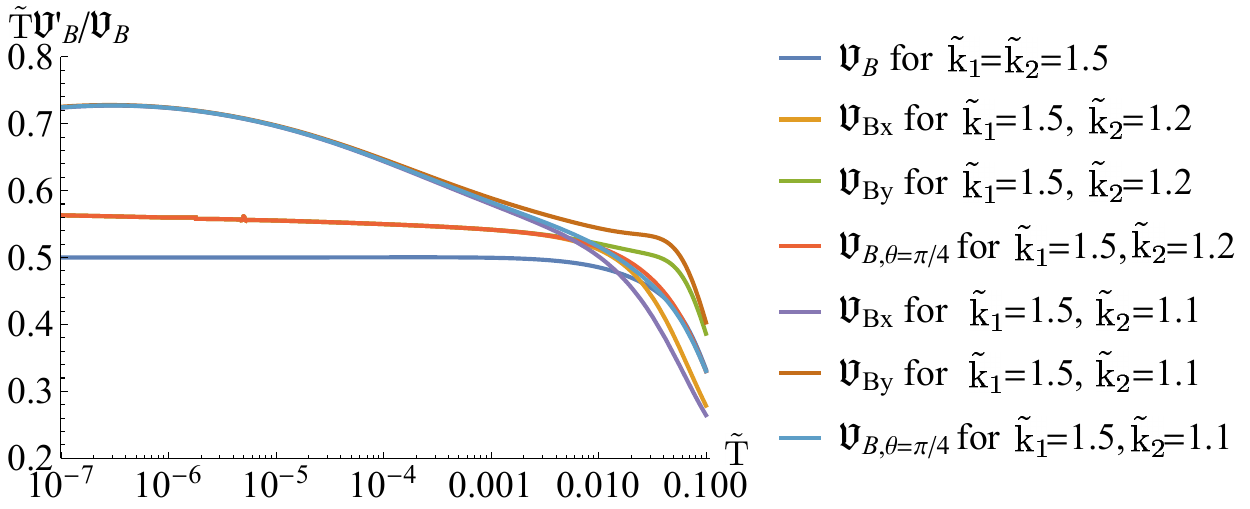}\ \\
  \caption{$\tilde T\mathfrak v_B'/\mathfrak v_B \,v.s.\, \tilde T$ over anisotropy background for $\tilde k_1=1.5$ and $\tilde k_2=1.5,\,1.2,\,1.1$, respectively. The system is  metallic phase in any direction.}
  \label{fig:vbvstvsanisov1}
\end{figure}
\begin{figure}[htbp]
  \centering
  \includegraphics[width=0.6\textwidth]{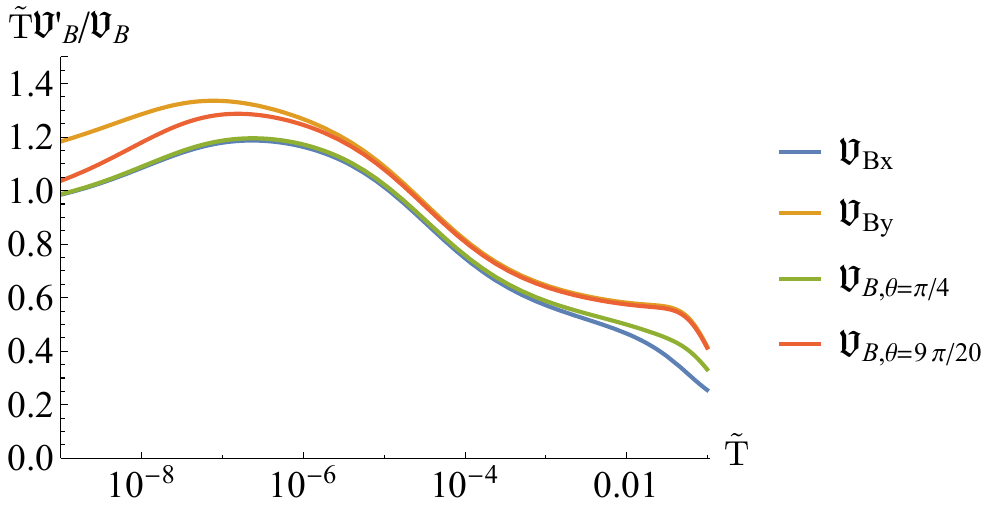}\ \\
  \caption{$\tilde T\mathfrak v_B'/\mathfrak v_B \,v.s.\, \tilde T$ over anisotropy background for $\tilde k_1=1.5$ and $\tilde k_2=1$. Notice that this system is still metallic phase in any direction.}
  \label{fig:vbvstvsanisov2}
\end{figure}

Then we study the anisotropy two-dimensional Q-lattice model. The first case is that the system is a metallic phase in any direction. Fig. \ref{fig:vbvstvsanisov1} shows $\tilde T\mathfrak v_B'/\mathfrak v_B \,v.s.\, \tilde T$ over anisotropy background for $\tilde k_1=1.5$ and $\tilde k_2=1.5,\,1.2,\,1.1$, respectively. First, we see that for fixed lattice parameter, $\tilde T\mathfrak v_B'/\mathfrak v_B$ tends to converge to a fixed value down to ultra-low temperature $\tilde T=10^{-7}$ for different angle $\theta$. This actually reflects the fact that $V_{xx}/V_{yy}$ is regular in zero-temperature limit when both directions are in metallic phase. The angular independence of the $\tilde T\mathfrak v_B'/\mathfrak v_B$ means that $\lim_{T\to 0}\beta \to 0$, from Eq. \eqref{eq:texpdv0} we find that the $\mathfrak v_B(\theta)$ converges to a fixed value along any direction. Secondly, for different lattice parameters, $\tilde T\mathfrak v_B'/\mathfrak v_B$ tends to converge to a different value in the zero-temperature limit, which indicates that the theory flows to different IR fixed points. This is a key difference from the case with a single lattice, where the metallic phases are shown related only to the AdS$_2 \times \mathbb R^2$ IR fixed points.

However, when the anisotropy becomes larger (here, we take $\tilde k_1=1.5$ and $\tilde k_2=1$), we observe an novel phenomenon that, at least down to the $\tilde T\sim 10^{-9}$, $v_B$ does no longer follow the behavior $v_B\sim \tilde T^{\alpha}$ (see Fig. \ref{fig:vbvstvsanisov2}). Notice that this system is still metallic phase in any direction. On the other hand, we also observe that the $\tilde Tv'_{Bx}/v_{Bx}$ is smaller than that of $\tilde Tv'_{By}/v_{By}$, while the $\tilde Tv'_{B,\theta=\pi/4}/v_{B,\theta=\pi/4}$ converges to that of $\tilde Tv'_{Bx}/v_{Bx}$. This suggests that in the limit of zero temperature, $\beta$ does not tend to vanishing and we conclude that $\beta>0$.

Another anisotropic case is that the system is in metallic phase in $x$ direction and insulating phase in $y$ direction. The behaviors for this case are shown in Fig. \ref{fig:vbvstvsanisov4}. We see that along $x$ direction, $\tilde T\mathfrak v_B'/\mathfrak v_B$ converges to a fixed value being smaller than unity. While along $y$ direction, $\tilde T\mathfrak v_B'/\mathfrak v_B$ converges to a fixed value being larger than unity. Notice that, the $\tilde T\mathfrak v'_{Bx}/\mathfrak v_{Bx}$ is smaller than that of $\tilde T\mathfrak v'_{By}/\mathfrak v_{By}$, while the $\tilde T\mathfrak v'_{B,\theta=\pi/4}/\mathfrak v_{B,\theta=\pi/4}$ seems converge to that of $\tilde T\mathfrak v'_{By}/\mathfrak v_{By}$. This situation is delicate, since $\tilde T\mathfrak v'_{Bx}/\mathfrak v_{Bx}<\mathfrak v'_{By}/\mathfrak v_{By}$ suggests $\beta>0$, while $\tilde T\mathfrak v'_{B,\theta=\pi/4}/\mathfrak v_{B,\theta=\pi/4}$ converging to $\tilde T\mathfrak v'_{Bx}/\mathfrak v_{Bx}$ suggests $\beta<0$. This seemingly contradicted conclusion is on account of the temperature so far is not low enough. Closer examination of the zoom in inset in Fig. \ref{fig:vbvstvsanisov4} will reveal that, $\tilde T\mathfrak v'_{B,\theta=\pi/4}/\mathfrak v_{B,\theta=\pi/4}$ starts to deviate from $\tilde T\mathfrak v'_{By}/\mathfrak v_{By}$, and it will converge to $\tilde T\mathfrak v'_{Bx}/\mathfrak v_{Bx}$ in even lower temperature. In this case, we still have that $\beta>0$.
\begin{figure}[htbp]
  \centering
  \includegraphics[width=0.6\textwidth]{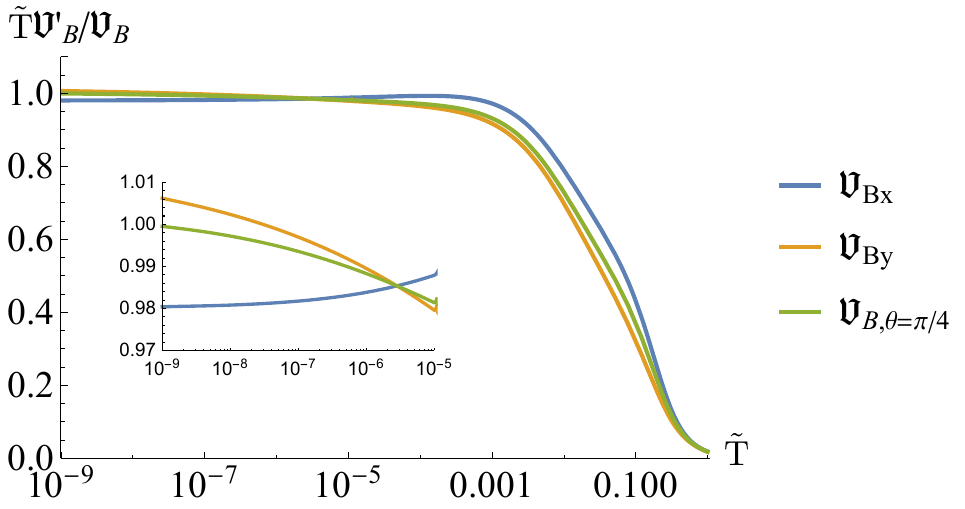}\ \\
  \caption{$\tilde T\mathfrak v_B'/\mathfrak v_B \,v.s.\, \tilde T$ over anisotropy background for $\tilde k_1=0.5$ and $\tilde k_2=0.1$. The system is in metallic phase in $x$ direction and insulating phase in $y$ direction.}
  \label{fig:vbvstvsanisov4}
\end{figure}

Next, we elaborate on the anisotropic case where the system is insulating phase in any direction (Fig. \ref{fig:vbvstvsanisov3}). $\tilde T\mathfrak v_B'/\mathfrak v_B$ converges to different fixed value equaling to or being smaller than unity for different angle $\theta$. The $\tilde T\mathfrak v'_{Bx}/\mathfrak v_{Bx}$ and $\tilde T\mathfrak v'_{By}/\mathfrak v_{By}$ converge to slightly different values, while $\tilde T\mathfrak v'_{B,\theta=\pi/4}/\mathfrak v_{B,\theta=\pi/4}$ converges to that of $\tilde T\mathfrak v'_{By}/\mathfrak v_{By}$. This means that $V_{xx}/V_{yy}\sim \tilde T^{\beta}$ with $\beta<0$, i.e., the $\lim_{T\to 0} V_{xx}/V_{yy} \to \infty$. This is different from the case where the system is in the metallic phase in any direction.

\begin{figure}[htbp]
  \centering
  \includegraphics[width=0.6\textwidth]{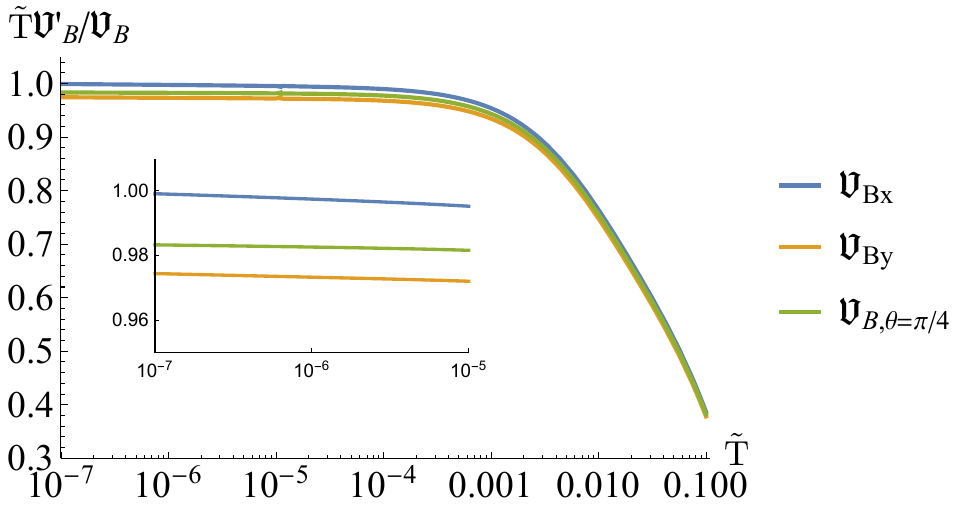}\ \\
  \caption{$\tilde T\mathfrak v_B'/\mathfrak v_B \,v.s.\, \tilde T$ over anisotropy background for $\tilde k_1=0.1$ and $\tilde k_2=0.01$. The system is insulating phase in any direction.}
  \label{fig:vbvstvsanisov3}
\end{figure}

Before closing this subsection, we would like to give some comments. First, according to the properties of the DC conductivity with the temperature, the anisotropic system can be categorized into three classes:
\begin{itemize}
  \item Case 1: The system is in the metallic phase in any direction.
  \item Case 2: The system is in the metallic phase in $x$ direction and insulating phase in $y$ direction.
  \item Case 3: The system is in the insulating phase in any direction.
\end{itemize}
Further, by using the scaling behavior of the butterfly, we can meticulously depict this anisotropic system by the parameter $\beta$.
\begin{itemize}
  \item When the system is in the metallic phase in any direction (case 1), $\beta=0$ for mildly anisotropic lattice.
  \item When further increasing the anisotropy, $\beta>0$. This phenomenon persists from case 1 to case 2.
  \item When the system is insulating phase in any direction, $\beta<0$.
\end{itemize}
To summarize, for anisotropic cases, the most important property is that the IR fixed point can be non-AdS$_2 \times \mathbb R^2$ even for metallic phases. Also, we observed some patterns of $\alpha$ and $\beta$ for different cases. This may help explain the bizarre IR fixed point. In addition, the above phenomena show that in most cases, there is indeed a scaling relationship between butterfly velocity and temperature. Moreover, the scaling relationship of butterfly velocity along different directions is indeed consistent with the deduced relationship Eq. \eqref{eq:texpd}. This shows that the above numerical results do verify the hypothesis given in Eq. \eqref{eq:texp} and Eq. \eqref{eq:texpd}.

\section{Charge diffusion}\label{sec:charge}

In this section, we study the charge diffusion constant $D_c$ in different phases. In \cite{Hartnoll:2014lpa}, it was conjectured that charge diffusion constant has a lower bound set by a relaxation time $\tau_P$
\fa
\label{diffusion-p}
D_c\geqslant v^2\tau_P\geqslant v^2\frac{\hbar}{k_B\tilde T}\,,
\ffa
with a characteristic velocity $v$. Here, we have used the following relation \cite{Zaanen:2004,Sachdev:2011cs}
\fa
\label{tau}
\tau_P\sim\frac{\hbar}{k_B\tilde T}\,.
\ffa
In what follows, we shall set $k_B=\hbar=1$.

Following the idea in \cite{Blake:2016wvh}, we identify the characteristic velocity $v$ in \eqref{diffusion-p} as the butterfly velocity $\mathfrak v_B$. Further, we assume that
\fa
\label{Dcvb}
D_c=\Omega\frac{\mathfrak v_B^2}{\tilde T}\,.
\ffa
We parameterize the above equation by $\Omega$ which depends on the details of the model. Notice that when the background is anisotropic, both the diffusion constant and the butterfly velocity depend on the angle $\theta$. The conjecture \eqref{diffusion-p} holds when $\Omega$ is a pure number of order $1$.
In a holographic theory with a particle-hole symmetry \cite{Blake:2016wvh,Blake:2016sud}, indeed it was found that there exists a universal regime in which $\Omega$ is a constant that only depends on the IR theory. However, such a universal regime cannot be found in the holographic theory at finite density \cite{Kim:2017dgz,Jeong:2017rxg}. Although we cannot find such a universal regime in the holographic theory at finite density, we expect that there is still a scaling behavior at extremely low temperature when the holographic theory is at the insulating phase. We shall numerically confirm this in what follows.

In the incoherent limit where strong momentum dissipation is present, the thermal current decouples from the charge current.
In this case, the charge diffusion is related to the DC conductivity and the charge susceptibility $\chi$ by the simple Einstein relation
\fa
\label{Dcchi}
D_c=\frac{\sigma_{DC}}{\chi}\,.
\ffa
By definition, the charge susceptibility $\chi$ is
\fa
\label{chidef}
\chi=\frac{(\partial\rho/\partial\mu)_{V,\tilde T}}{\rho^2}\,.
\ffa
Since in the holographic theory, we have the scaling invariance, it is easy to derive the expression of the charge susceptibility $\chi$. First, the $\rho$ scales as
\fa
\label{rhoscale}
\rho\sim\mu^2\,.
\ffa
Then we must have
\fa
\label{rhoscaled}
\rho=\xi(\tilde k_{1,2},\tilde\lambda_{1,2},\tilde T)\mu^2\,,
\ffa
where $\xi$ is a constant depending on $\tilde k_{1,2}$, $\tilde\lambda_{1,2}$ and $\tilde T$. Therefore, we yield
\fa
\label{partialrhovsmu}
\frac{\partial\rho}{\partial\mu}=2\xi\mu\,.
\ffa
Substituting Eqs.\eqref{partialrhovsmu} and \eqref{rhoscaled} into the definition of the charge susceptibility (Eq. \eqref{chidef}), we have
\fa
\label{chif}
\chi=\frac{2}{\xi\mu^3}\,.
\ffa
Then the dimensionless charge susceptibility is
\fa
\label{chidim}
\tilde{\chi}=\chi\mu^3=\frac{2\mu^2}{\rho}\,.
\ffa
Therefore, the dimensionless charge diffusion is
\fa
\label{DCdim}
\tilde{D}_c=\frac{\sigma_{DC}}{\tilde{\chi}}\,.
\ffa

\begin{figure}[htbp]
  \centering
  \includegraphics[width=0.7\textwidth]{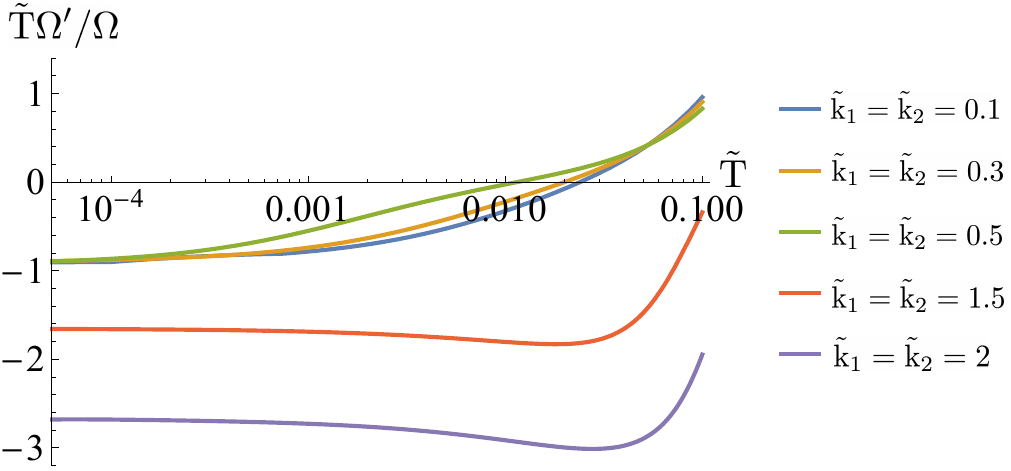}\ \\
  \caption{$\tilde T\Omega'/\Omega \,v.s.\, \tilde T$ for different phases over isotropy background ($\tilde k_1=\tilde k_2=1.5,\,2$ corresponds
    to metallic phases and $\tilde k_1=\tilde k_2=0.1,\,0.3,\,0.5$ corresponds to insulating phases).
  }
  \label{fig:disotropy}
\end{figure}

By Eqs.\eqref{Dcchi} and \eqref{DCdim}, we can numerically work out the parameter $\Omega$ as the function of $\tilde T$ in different phases. 
When $k_{1,2}$ increases, the momentum dissipation is enhanced. We calculate the $\tilde T\Omega'/\Omega$ at different $k$ values, and deduce the phenomenon of incoherent limit by increasing $k_{1,2}$.
Fig. \ref{fig:disotropy} shows $\tilde T\Omega'/\Omega$ as the function of $\tilde T$ for different phases over isotropic backgrounds, where the butterfly velocity, as well as the diffusion constant, are all isotropic. 
We see that for the insulating phase $\tilde T\Omega'/\Omega$ tends to a fixed value at the extremal low temperature. It implies that the parameter $\Omega$ follows the scaling behavior as $\Omega\sim \tilde T^{\gamma}$ with $\gamma<0$. But for the metallic phase, $\tilde T\Omega'/\Omega$ does not follow the same behavior at the low temperature. Notice that in isotropic cases, the exponent of $\mathfrak v_B$ with temperature is $1 / 2$, but $\Omega$ does not have the same stable exponential relationship. From the definition, the low-temperature behavior of $\Omega$ is related to the low-temperature behavior of $v_B$, $\sigma_{DC}$ and $\rho$. At present, the scaling behavior of $\sigma_{DC}$ with temperature is still unclear. Moreover, $\rho$ is different from $v_B$ and $\sigma_{DC}$ in that it is not dictated by the horizon data, on the contrary, it is related to the whole background solution. Therefore, at present, the analytical understanding of this behavior remains to be explored. 

 More importantly, the scaling behavior of $\Omega\sim \tilde T^{\gamma}$ suggests that the conjecture \eqref{diffusion-p} does not hold in the current model. This means that the characteristic relaxation time scale in this system does not saturates the bound set by the Planckian time scale. Here, we obtained $\gamma<0$, which means that the relaxation time $\tau\sim \tilde T^{\gamma-1}$ is much larger than the Planckian time scale at low temperatures. Notice that there is also a large relaxation time in Fermi liquid theory as $\tau \sim T^{-2}$. Although the strongly correlated system is beyond the Fermi liquid theory, it is still possible to obtain a large relaxation time similar to that in Fermi liquid. This may imply that, although we have obtained many supportive examples for the saturation of universal bounds, the strong correlation theory does not always push up characteristic parameters against and saturate any bound.

\begin{figure}[htbp]
  \centering
  \includegraphics[width=0.45\textwidth]{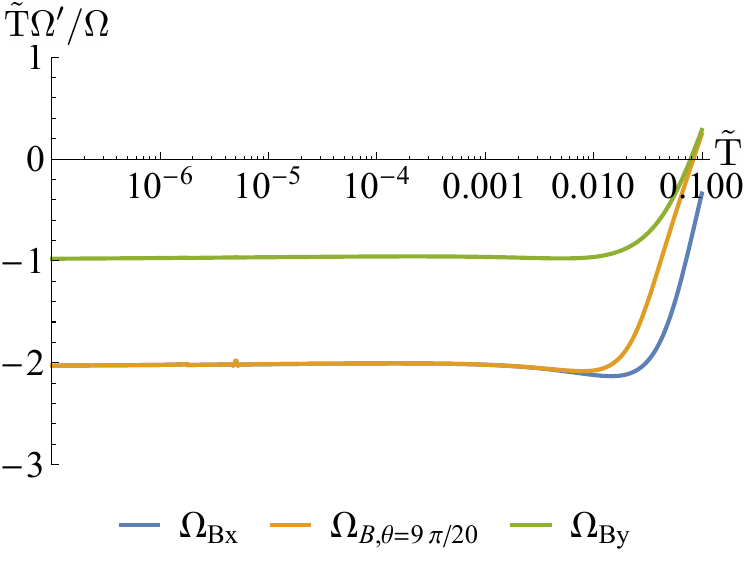}\ \hspace{0.5cm}
  \includegraphics[width=0.45\textwidth]{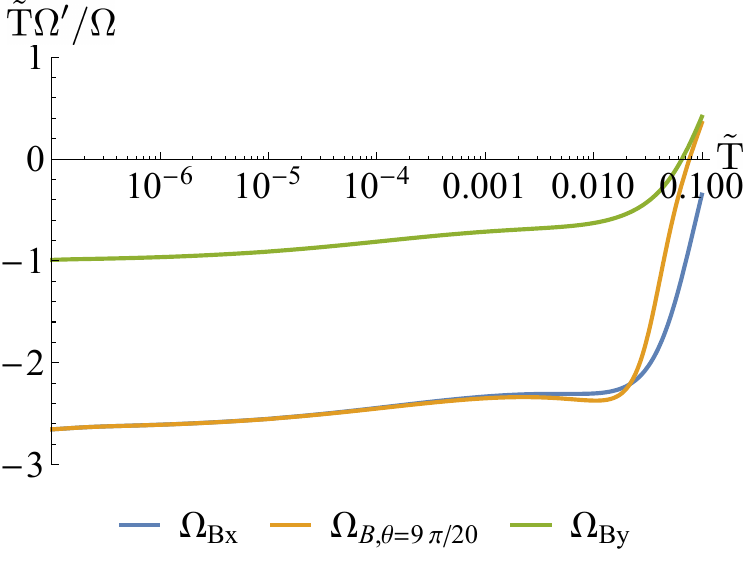}
  \caption{$\tilde T\Omega'/\Omega \,v.s.\, \tilde T$ over anisotropy background. Left plot is for $\tilde k_1=1.5$ and $\tilde k_2=1.2$ and right plot is for $\tilde k_1=1.5$ and $\tilde k_2=1.1$. The system is  metallic phase in any direction.}
  \label{fig:danisov1}
\end{figure}
\begin{figure}[htbp]
  \centering
  \includegraphics[width=0.7\textwidth]{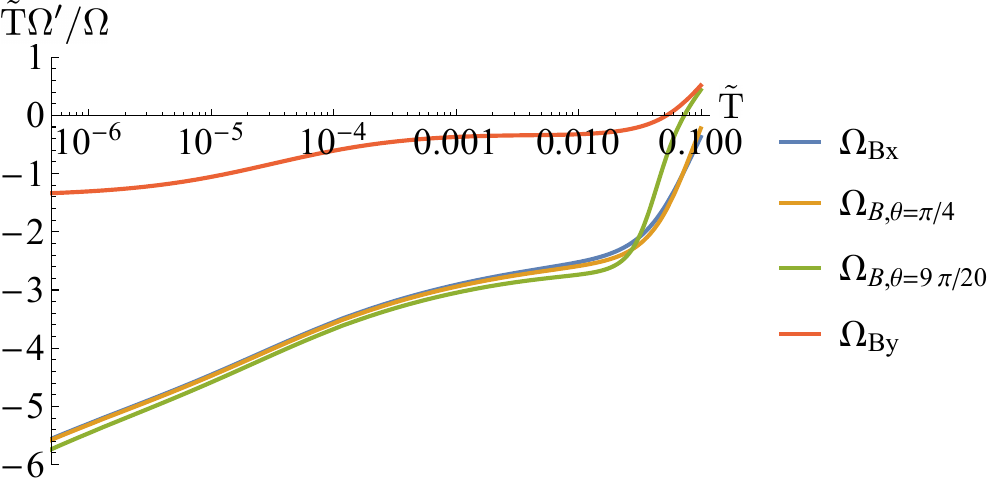}\ \\
  \caption{$\tilde T\Omega'/\Omega \,v.s.\, \tilde T$ over anisotropy background for $\tilde k_1=1.5$ and $\tilde k_2=1$. Notice that this system is still metallic phase in any direction.}
  \label{fig:danisov2}
\end{figure}
\begin{figure}[htbp]
  \centering
  \includegraphics[width=0.45\textwidth]{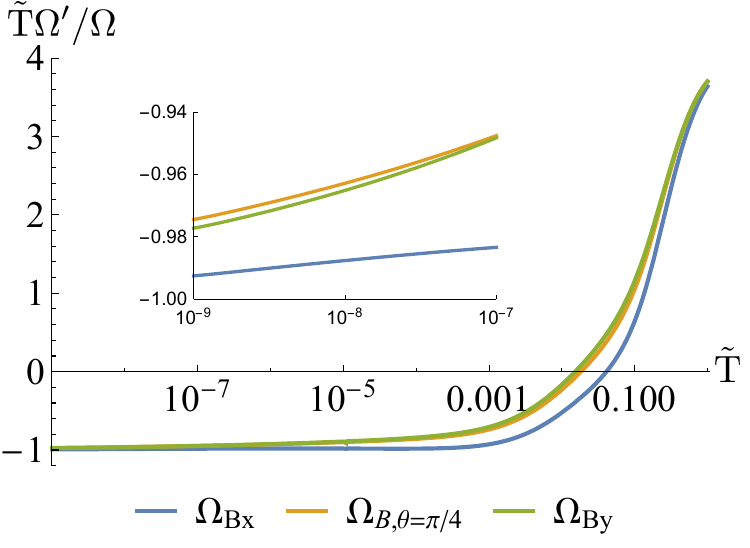}\ \hspace{0.5cm}
  \includegraphics[width=0.45\textwidth]{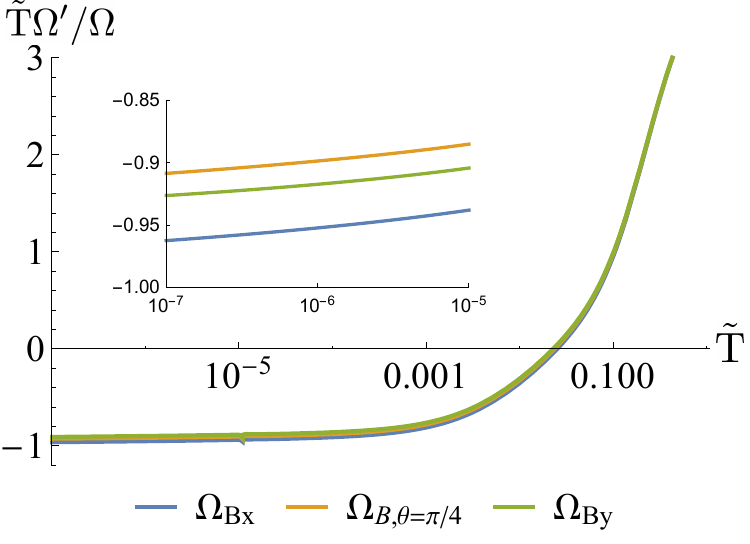}
  \caption{$\tilde T\Omega'/\Omega \,v.s.\, \tilde T$ over anisotropy background. Left plot is for $\tilde k_1=0.5$ and $\tilde k_2=0.1$, for which the system is in metallic phase in $x$ direction and insulating phase in $y$ direction. Right plot is for $\tilde k_1=0.1$ and $\tilde k_2=0.01$, for which the system is insulating phase in any direction.}
  \label{fig:danisov3}
\end{figure}

Also, we study the behavior of the parameter $\Omega$ over the anisotropy background
(Fig. \ref{fig:danisov1}, Fig. \ref{fig:danisov2} and Fig. \ref{fig:danisov3}). We summarize the main characteristics as what follows:
\begin{itemize}
  \item When the system is in metallic phase in any direction and the anisotropy is mild (right plot in Fig. \ref{fig:danisov1} for $\tilde k_1=1.5$ and $\tilde k_2=1.1$), the curves of $\tilde T\Omega'/\Omega \,v.s.\, \tilde T$ for $\theta=\pi/2$ and all other angles except $\theta=\pi/2$ flow to a fixed value at extremal low temperature, respectively. But when the anisotropy becomes larger (right plot in Fig. \ref{fig:danisov2} for $\tilde k_1=1.5$ and $\tilde k_2=1.2$), although the curves of $\tilde T\Omega'/\Omega \,v.s.\, \tilde T$ are consistent at low temperature for all angles except the one of $\theta=\pi/2$ as that for $\tilde k_1=1.5$ and $\tilde k_2=1.2$, $\tilde T\Omega'/\Omega \,v.s.\, \tilde T$ does not converge to a certain fixed value. Especially, as the anisotropy further becomes larger (Fig. \ref{fig:danisov2} for $\tilde k_1=1.5$ and $\tilde k_2=1$), this phenomenon becomes more evident.
  \item For systems in metallic phase in $x$ direction but in insulating phase in $y$ direction, and systems in insulating phase in any direction, the converging value of $\tilde T\Omega'/\Omega \,v.s.\, \tilde T$ will vary with angle.
\end{itemize}

From the above behavior, we can see that $\Omega$ presents richer properties in the case of anisotropy. When the anisotropy is weak, the $\Omega$ tends to two distinct scaling relations at low temperature. When the anisotropy is enhanced, the low-temperature scaling of $\Omega$ in different directions shows rich scaling behavior, depending on the specific angle. The direct reason for this phenomenon is the rich behavior of $\mathfrak v_B$ itself with the angle. Meanwhile, it is worth noting that $\rho$ is related to the whole background solution, thus the analytical understanding of these phenomena needs to be further explored.

\section{Discussion}\label{sec:discussion}

In this paper, we have studied the anisotropy effect on the dynamical quantities, including the DC conductivity, the butterfly velocity and the charge diffusion. The spatial anisotropy is realized in Q-lattice model by setting the wavenumber $\tilde k_1\neq \tilde k_2$. We reveal the phase structure by DC conductivity. The anisotropy lays a significant effect on the phase structures, that we divide into three classes: metallic, metallic and insulating in different directions, insulating.

When the anisotropy is introduced, MIT is still present. But MIT phase diagram along $x$ direction is greatly affected by the lattice parameters along $y$ direction. An interesting result is that $\tilde k_2=0$ sets the upper bound of the critical line in the phase diagram ($\tilde\lambda_1$, $\tilde k_1$). As $\tilde k_2$ increases, this critical line goes up and approaches this upper bound. That is to say, the phase structure is novel that one lattice can affect the other lattice. We provided an analytical understanding through mode analysis and explained this phenomenon from the perspective of dual physics.

The butterfly velocity is also angle-dependent when the anisotropy is introduced. We further confirm the result that the lattice suppresses the butterfly velocity. It has been observed in our previous work \cite{Ling:2016ibq}, in which the lattice along $y$ direction is turned off. In addition, we also confirm the robustness of the butterfly velocity as the diagnostic of the QPT in an anisotropic system.

The charge diffusion still diagnoses the MIT, but shows different properties from the butterfly velocity. First, in isotropic cases, $\Omega$ tends to different low-temperature behavior in metallic phases; while in insulating phases, it tends to the same low-temperature behavior. The reason for this phenomenon is that the charge density is not dictated by the horizon data, and the low-temperature scaling behavior of DC conductivity is still unclear. In addition, the charge diffusion also shows rich behavior in the case of anisotropy: when the anisotropy is weak, $\Omega$ tends to two distinct values, while when the anisotropy is strong, the low-temperature behavior of $\Omega$ is obviously related to the angle. Meanwhile, the charge diffusion behavior in this model also shows that the relaxation time of the system does not saturate the bound set by the Planckian time scale, but is much larger than it.

In order to fully understand the effect of anisotropy, we need to further study the IR fixed points and the scaling behavior of charge density for the model in consideration. Moreover, the dynamic properties of anisotropic ion lattice are also worth exploring. At this time, solving the background solutions needs more complicated techniques for partial differential equations. In addition, the quantum information measurement in this anisotropic background, such as holographic entanglement entropy and the entanglement wedge minimum cross-section, is also worth studying. We will further explore these directions in our recent work.

\section*{Acknowledgments}
Peng Liu would like to thank Yun-Ha Zha for her kind encouragement during this work. This work is supported by the Natural Science Foundation of China under Grant No. 11905083, 11775036, 12147209. J. P. Wu is also supported by Top Talent Support Program from Yangzhou University.

\end{document}